\documentclass[5pt]{article}
\usepackage{amsmath}
\usepackage[latin1]{inputenc}
\usepackage{amsmath,amsthm,amssymb}
\usepackage{graphicx}
\usepackage{color}

\renewcommand{\theequation}{\thesection.\arabic{equation}}

\newtheorem{lemma}{Lemma}[section]
\newtheorem{thm}{Theorem} [section]

\newtheorem{exmp}{Example} [section]

\newtheorem{rem}{Remark}[section]
\title{A New Constructions of Minimal Binary Linear Codes
\thanks {The work of H.B. Liu is supported by the Natural Science Foundation of China with No.11901062, the work of Q.Y. Liao is supported by the Natural Science Foundation of China with No.12071321.}}
\author{{Haibo Liu\footnote{\ H.B.Liu, School of Applied Mathematics, Chengdu University of Information Technology,
 \small Chengdu, Sichuan, China (Email:liuhaibo@cuit.edu.cn)}
\quad Qunying Liao\footnote{\ Q.Y.Liao, School of Mathematical Sciences, Sichuan Normal University,
 \small Chengdu, Sichuan, China (Email:qunyingliao@sicnu.edu.cn)}
}\ }
\date{}
\begin{document}
\baselineskip15pt \maketitle
\renewcommand{\theequation}{\arabic{section}.\arabic{equation}}
\catcode`@=11 \@addtoreset{equation}{section} \catcode`@=12
\begin{abstract}

Recently, minimal linear codes have been extensively studied due to their applications in secret sharing schemes, secure two-party computations, and so on. Constructing minimal linear codes violating the Ashikhmin-Barg condition and then determining their weight distributions have been interesting in coding theory and cryptography. In this paper, a generic construction for binary linear codes with dimension $m+2$ is presented, then a necessary and sufficient condition for this binary linear code to be minimal is derived. Based on this condition and exponential sums, a new class of minimal binary linear codes violating the Ashikhmin-Barg condition is obtained, and then their weight enumerators are determined.


 \end{abstract}

 \section{Introduction and Background}
 $ $

Linear codes over finite fields have been extensively studied by researchers according to specific applications in computer and communication systems, data storage devices, and consumer electronics. Minimal linear codes as a special class of linear codes have been widely applied in secret sharing schemes and secure two-party computations \cite{Carlet-Ding-Yuan}\cite{Yuan-Ding}, and so on.

Minimal codewords can be used in linear codes based-access structures in secret sharing schemes, which are protocols with a distribution algorithm, implemented by a dealer and some participants, see \cite{Shamir}. The dealer splits a secret $S$ into different pieces and then distributes them to participants set $\mathcal{P}$, only authorized subsets of $\mathcal{P}$ (access structure $\Gamma$) can be able to reconstruct the secret by using their respective shares. A set of participants $D$ is called a minimal authorized subset if $D\in\Gamma$ and no proper subset of $D$ belong to $\Gamma$. In \cite{Massey}, Massey pointed out that the support of minimal codewords of the dual code can give the access structure of a secret sharing scheme. But to describe the set of minimal codewords of a linear code is quite difficult in general, even in the binary case. To simplify this task, one can try to find linear codes with any codeword minimal, called minimal linear codes. The problem of finding minimal linear codes has first been investigated in \cite{Ding-Yuan}. A sufficient condition for a linear code to be minimal is given in the following lemma.
\begin{lemma}\cite{Ashikhmin-Barg}(Ashikhmin-Barg)\label{lem9}
Let $p$ be a prime, then a linear code $\mathcal{C}$ over $\mathbb{F}_p$ is minimal if
\[\frac{w_{min}}{w_{max}}>\frac{p-1}{p},\]
where $w_{min}$ and $w_{max}$ denote the minimum and maximum nonzero Hamming weights for $\mathcal{C}$, respectively.
\end{lemma}
With the help of Lemma \ref{lem9}, many minimal linear codes were constructed from linear codes with a few weights \cite{Ding1} \cite{Ding2} \cite{Ding-Ding} \cite{Tang-Li-Qi-Zhou-Helleseth} \cite{Xiang}. The sufficient condition in Lemma \ref{lem9} is not usually necessary for a linear code to be minimal. In recent years, searching for minimal linear codes with $\frac{w_{min}}{w_{max}}\leq \frac{p-1}{p}$ has been an interesting research topic. In 2018, Chang and Hyun \cite{Chang-Hyun} made a breakthrough and constructed an infinite family of minimal binary linear codes with $\frac{w_{min}}{w_{max}}< \frac{1}{2}$ by the generic construction
\begin{eqnarray}\label{shizi1}
\mathcal{C}_f=\{(uf(\mathbf{x})+\mathbf{v}\cdot \mathbf{x})_{\mathbf{x}\in {\mathbb{F}_p^m}^\ast}:u\in \mathbb{F}_p,\mathbf{v}\in\mathbb{F}_p^m\}.
\end{eqnarray}

Based on the generic construction, a lot of minimal linear codes are obtained, which do not satisfy the Ashikhmin-Barg condition. Ding et al. \cite{Ding-Heng-Zhou} gave a necessary and sufficient condition for a binary linear code to be minimal, and then employed special Boolean functions to obtain three classes of minimal binary linear codes. Heng et al. \cite{Heng-Ding-Zhou} used the characteristic function of a subset in $\mathbb{F}_3^m$ to construct a class of minimal ternary linear codes. Bartoli and Bonini \cite{Bartoli-Bonini} generalized the construction of minimal linear codes in \cite{Heng-Ding-Zhou} from ternary case to be odd characteristic case. Bonini and Borello \cite{Bonini-Borello} presented a family of minimal codes arising from cutting blocking sets. Tao et al. \cite{Tao-Feng-Li} obtained three-weight or four-weight minimal linear codes by using partial difference sets. Meanwhile, there are other ways to construct minimal linear codes violating the Ashikhmin-Barg condition, see  \cite{Bartoli-Bonini-Gunes} \cite{Li-Yue} \cite{Liu-Liao} \cite{Lu-Wu-Cao} \cite{Tang-Qi-Liao-Zhou} \cite{Xu-Qu}.

Till now, a lot of minimal linear codes violating the Ashikhmin-Barg condition are constructed, and their weight distributions are given, but the dimension of these minimal linear codes is either $m$ or $m+1$, which leads the code rate lower.

In this paper, we give a generic construction for minimal binary linear codes violating the Ashikhmin-Barg condition with dimension $m+2$. The paper is organized as follows. Section 2 provides some definitions and notations, which will be needed in the sequel. Section 3 presents a generic construction for binary linear codes with dimension $m+2$, and gives a necessary and sufficient condition for this binary linear codes to be minimal. Basing on this condition and exponential sums, Section 4 obtains a new class of minimal binary linear codes violating the Ashikhmin-Barg condition from partial spread, and then determines their weight enumerators. Section 5 concludes the whole paper and gives the further study.

\section{Preliminaries}
$ $

In this section we give some definitions and notations for minimal linear codes. Throughout the whole paper, let $p$ be a prime, denote $\mathbb{F}_p$ to be the finite field with $p$ elements. An $[n,k,d]$ linear code $\mathcal{C}$ over $\mathbb{F}_ p$ is a $k$-dimensional subspace of $\mathbb{F}_p^n$ with minimum (Hamming) distance $d$. For any $i=1,\dots,n$, $A_i$ denotes the number of codewords with Hamming weight $i$ in $\mathcal{C}$ of length $n$. The $\emph{weight}$ $\emph{enummerator}$ of $\mathcal{C}$ is defined by
\[1+A_1z+\cdots+A_nz^n.\]
The $\emph{weight}$ $\emph{distribution}$ $(1,A_1,\dots,A_n)$ is important in coding theory, since it contains some crucial information as to estimate the error-correcting capability and the probability of error-detection and correction with respect to some algorithms \cite{Ding-Ding}. $\mathcal{C}$ is said to be a $t$-weight code if the number of nonzero $A_i$ in the sequence $(A_1,\dots,A_n)$ is equal to $t$.

For a vector $\mathbf{a}=(a_1,\dots,a_m)\in \mathbb{F}_p^m$, the $\emph{S}upport$ of $\mathbf{a}$ is defined by
\[\text{Supp}(\mathbf{a})=\{1\leq i\leq m:a_i\neq 0\}.\]
Let $wt(\mathbf{a})$ be the Hamming weight of $\mathbf{a}$, then $wt(\mathbf{a})=|\text{Supp}(\mathbf{a})|$. For $\mathbf{b}\in \mathbb{F}_p^m$, we say that $\mathbf{a}$ covers $\mathbf{b}$ if $\text{Supp}(\mathbf{b})\subseteq \text{Supp}(\mathbf{a})$, denoted by $\mathbf{b}\preceq \mathbf{a}$. A codeword $\mathbf{c}$ in $\mathcal{C}$ is minimal if $\mathbf{c}$ covers only those codewords $u\mathbf{c}$ $(u\in \mathbb{F}_p^\ast)$. $C$ is said to be minimal if every codeword in $\mathcal{C}$ is minimal.

A function $f$ from $\mathbb{F}_2^m$ to $\mathbb{F}_2$ is called a Boolean function also. For a Boolean function $f$, the Walsh transform is defined by
\[\widehat{f}(\mathbf{w})=\sum_{\mathbf{x}\in \mathbb{F}_2^m}(-1)^{f(\mathbf{x})+\mathbf{w}\cdot \mathbf{x}},\]
where $\mathbf{w}\in \mathbb{F}_2^m$ and $\mathbf{w}\cdot \mathbf{x} $ is the standard inner product of $\mathbf{w}$ and $\mathbf{x}$. Another related Walsh transform of $f$ is defined by
\[\widetilde{f}(\mathbf{w})=\sum_{\mathbf{x}\in \mathbb{F}_2^m}f(\mathbf{x})(-1)^{\mathbf{w}\cdot \mathbf{x}},\]
where $f(x)$ is viewed as a real-valued function taking on only $0$ or $1$. The relation between two kinds of Walsh transforms is well-known and listed below.
\begin{lemma}\label{lem3}
Let $f(x)$ be a Boolean function, then
 \begin{eqnarray*}
\widehat{f}(\mathbf{w})=\left\{\begin{array}{ll}2^m-2\widetilde{f}(\mathbf{0}), &\text{if}\quad w=\mathbf{0};\\
 -2\widetilde{f}(\mathbf{w}), &\text{if}\quad  \mathbf{w}\neq \mathbf{0}. \end{array}\right.
\end{eqnarray*}
\end{lemma}
If $\mathbf{b}\preceq \mathbf{a}$, the relation between $wt(\mathbf{a})$ and $wt(\mathbf{b})$ is given below.
\begin{lemma}\cite{Ding-Heng-Zhou}\label{lem1}
Let $\mathbf{a}\in \mathbb{F}_2^m$ and $\mathbf{b}\in\mathbb{F}_2^m$, then $\mathbf{b}\preceq \mathbf{a}$ if and only if
\[ wt(\mathbf{a}+\mathbf{b})=wt(\mathbf{a})-wt(\mathbf{b}).\]
\end{lemma}

\section{A General Construction of Minimal Binary Linear Codes from Boolean Functions}

$ $

In this section, we introduce a generic construction of binary codes from Boolean functions. Let $f(x)$ and $g(x)$ be Boolean functions from $\mathbb{F}_2^m$ to $\mathbb{F}_2$ such that the following conditions hold,

$\bullet$ $f(\mathbf{0})=0$ and $f(\mathbf{b})=1$ for at least one $\mathbf{b}\in \mathbb{F}_2^m$;

$\bullet$ $g(\mathbf{0})=0$ and $g(\mathbf{a})=1$ for at least one $\mathbf{a}\in \mathbb{F}_2^m$;

$\bullet$ $f(\mathbf{e})+g(\mathbf{e})=1$ for at least one $\mathbf{e}\in \mathbb{F}_2^m$.

We now define a linear code by
\begin{eqnarray}\label{shizi2}
\mathcal{C}_{f,g}=\left\{\mathbf{c}_{f,g}(\mathbf{v})=(uf(\mathbf{x})+rg(\mathbf{x})+\mathbf{v}\cdot \mathbf{x})_{\mathbf{x}\in {\mathbb{F}_2^m}^\ast}:u\in \mathbb{F}_2, r\in\mathbb{F}_2, \mathbf{v}\in \mathbb{F}_2^m\right\}.
\end{eqnarray}
Note that for $p=2$, the code $\mathcal{C}_f$ in (\ref{shizi1}) is a subcode of $\mathcal{C}_{f,g}$.  The dimension and weight distribution of $\mathcal{C}_{f,g}$ is given by the following theorem.
\begin{thm}\label{thm1}
The binary code $\mathcal{C}_{f,g}$ in (\ref{shizi2}) has length $2^m-1$ and dimension $m+2$ if  $f(\mathbf{x})\neq \mathbf{a} \cdot \mathbf{x}$, $g(\mathbf{x})\neq \mathbf{b} \cdot \mathbf{x}$ and $f(\mathbf{x})+g(\mathbf{x})\neq \mathbf{e} \cdot \mathbf{x}$ for all $\mathbf{a}, \mathbf{b} ,\mathbf{e}\in \mathbb{F}_2^m$. In addition, the weight distribution of $\mathcal{C}_{f,g}$ is given by the following multiset union:
\[\{\frac{2^m-\widehat{f}(\mathbf{w})}{2}: \mathbf{w}\in \mathbb{F}_2^m\}\cup\{\frac{2^m-\widehat{g}(\mathbf{w})}{2}: \mathbf{w}\in \mathbb{F}_2^m\}\cup\{\frac{2^m-\widehat{f+g}(\mathbf{w})}{2}: \mathbf{w}\in \mathbb{F}_2^m\}
\cup\{2^{m-1}:\mathbf{w}\in {\mathbb{F}_2^m}^\ast\}\cup\{0\}.\]
\end{thm}
$\mathbf{Proof.}$ We consider the weight distribution of $\mathcal{C}_{f,g}$ according to both $u=0$ or not and $r=0$ or not. We only give the proof of the case $u=r=1$, and omit the proofs of other cases, which are similar to the case $u=r=1$.

For $u=r=1$, we have
\begin{eqnarray*}
\widehat{f+g}(\mathbf{w})&=&\sum_{\mathbf{x}\in \mathbb{F}_2^m}(-1)^{f(\mathbf{x})+g(\mathbf{x})+\mathbf{w}\cdot \mathbf{x}}\\
&=&2^m-2|\{\mathbf{x}\in{\mathbb{F}_2^m}^\ast: f(\mathbf{x})+g(\mathbf{x})+\mathbf{w}\cdot\mathbf{x}=1\}|\\
&=&2^m-2wt(\mathbf{c}_{f,g}(\mathbf{w})).
\end{eqnarray*}
Thus, the weight distribution of $\mathcal{C}_{f,g}$ is obtained. Since each of $f$, $g$ or $f+g$ is not linear, the dimension of the code $\mathcal{C}_{f,g}$ must be $m+2$.

This completes the proof of Theorem \ref{thm1}. \hfill$\Box$\\

A natural question is when the code $\mathcal{C}_{f,g}$ in (\ref{shizi2}) is minimal. The next theorem gives a necessary and sufficient condition for $\mathcal{C}_{f,g}$ minimal.

\begin{thm}\label{thm2}
Let $\mathcal{C}_{f,g}$ be the code in Theorem \ref{thm1}, denote $F=\{f, g, f+g\}$, then $\mathcal{C}_{f,g}$ is minimal if and only if the following two conditions hold simultaneously for any $\mathbf{h}, \textbf{l}\in \mathbb{F}_2^m$.

(1) For any $f_1,f_2\in F$ and $\mathbf{h}\neq \textbf{l}$, we have
$$\widehat{f}_1(\mathbf{h})+\widehat{f}_2(\textbf{l})\neq 2^m \quad \text{and} \quad \widehat{f}_1(\mathbf{h})-\widehat{f}_2(\textbf{l})\neq 2^m;$$

(2) for any $f_1,f_2\in F$ with $f_1\neq f_2$, we have
$$\widehat{f}_1(\textbf{l}+\mathbf{h})+\widehat{f}_2(\textbf{l})-\widehat{f_1+f_2}(\mathbf{h})\neq 2^m\quad \text{and} \quad  \widehat{f}_1(\textbf{l}+\mathbf{h})+\widehat{f}_2(\mathbf{h})-\widehat{f_1+f_2}(\textbf{l})\neq 2^m.$$
\end{thm}
$\mathbf{Proof.}$ We define the linear code $S_m=\{\mathbf{s}_v=(\mathbf{v}\cdot \mathbf{x})_{\mathbf{x}\in {\mathbb{F}_2^m}^\ast}: \mathbf{v\in \mathbb{F}_2^m}\}$. Easily, $S_m$ is a Simplex code with parameters $[2^m-1,m,2^{m-1}]$. Clearly, each nonzero codeword in $S_m$ has weight $2^{m-1}$.

Let the vectors $\mathbf{f}$ and $\mathbf{g}$ be defined respectively as
\[\mathbf{f}=(f(\mathbf{x}))_{\mathbf{x}\in {\mathbb{F}_2^m}^\ast},\quad \text{and}\quad \mathbf{g}=(g(\mathbf{x}))_{\mathbf{x}\in {\mathbb{F}_2^m}^\ast}. \]
By definition, each codeword $\mathbf{c}_{f,g}(\mathbf{a})\in \mathcal{C}_{f,g}$ can be expressed as
\[\mathbf{c}_{f,g}(\mathbf{a})=u\mathbf{f}+r\mathbf{g}+\mathbf{s}_a,\]
where $u,r\in \{0,1\}$ and $\mathbf{s}_a\in S_m $. Then from the proof of Theorem \ref{thm1}, we can get
\begin{eqnarray}\label{shizi3}
wt(\mathbf{f}+\mathbf{s}_v )=2^{m-1}-\frac{\widehat{f}(\mathbf{v})}{2},
\end{eqnarray}
which will be employed for the proof of this theorem. We next consider the coverage of codewords in $\mathcal{C}_{f,g}$ by distinguishing the following two cases.

{\bf Case 1.} Two different codewords $\mathbf{c}_1$ and $\mathbf{c}_2$ come from the same type $\{\mathbf{s}_a,\mathbf{f}+\mathbf{s}_a,\mathbf{g}+\mathbf{s}_a,\mathbf{f}+\mathbf{g}+\mathbf{s}_a\}$. Then we have the following two subcases.

(1) For $\mathbf{c}_i=\mathbf{s}_a$ and $\mathbf{c}_j \in \{\mathbf{f}+\mathbf{s}_a,\mathbf{g}+\mathbf{s}_a,\mathbf{f}+\mathbf{g}+\mathbf{s}_a\}$, where $\{i,j\}=\{1,2\}$, we only give proof of $\mathbf{c}_1=\mathbf{s}_a$ and $\mathbf{c}_2=\mathbf{f}+\mathbf{g}+\mathbf{s}_a$, omit the proofs of other cases, which are similar to the case $\mathbf{c}_1=\mathbf{s}_a$ and $\mathbf{c}_2=\mathbf{f}+\mathbf{g}+\mathbf{s}_a$. By Lemma \ref{lem1} and (\ref{shizi3}), one can get
\[\mathbf{c}_1\preceq \mathbf{c}_2\Longleftrightarrow wt(\mathbf{f}+\mathbf{g})=wt(\mathbf{f}+\mathbf{g}+\mathbf{s}_a)-2^{m-1}\Longleftrightarrow \widehat{f+g}(\mathbf{0})-\widehat{f+g}(\mathbf{a})=2^m, \]
and
\[\mathbf{c}_2\preceq \mathbf{c}_1\Longleftrightarrow wt(\mathbf{f}+\mathbf{g})=2^{m-1}-wt(\mathbf{f}+\mathbf{g}+\mathbf{s}_a)\Longleftrightarrow\widehat{f+g}(\mathbf{0})+\widehat{f+g}(\mathbf{a})=2^m .\]

(2) Both $\mathbf{c}_1$ and $\mathbf{c}_2 $ belong to $ \{\mathbf{f}+\mathbf{s}_a,\mathbf{g}+\mathbf{s}_a,\mathbf{f}+\mathbf{g}+\mathbf{s}_a\}$ with $\mathbf{c}_1\neq \mathbf{c}_2$, we only give the proof of $\mathbf{c}_1=\mathbf{f}+\mathbf{s}_a$ and $\mathbf{c}_2=\mathbf{g}+\mathbf{s}_a$, omit the proofs of other cases, whose proofs are similar to this case. By Lemma \ref{lem1} and (\ref{shizi3}), we have
\[\mathbf{c}_1\preceq \mathbf{c}_2\Longleftrightarrow wt(\mathbf{f}+\mathbf{g})=wt(\mathbf{g}+\mathbf{s}_a)-wt(\mathbf{f}+\mathbf{s}_a)\Longleftrightarrow \widehat{f+g}(\mathbf{0})+\widehat{f}(\mathbf{a})-\widehat{g}(\mathbf{a})=2^m, \]
and
\[\mathbf{c}_2\preceq \mathbf{c}_1\Longleftrightarrow wt(\mathbf{f}+\mathbf{g})=wt(\mathbf{f}+\mathbf{s}_a)-wt(\mathbf{g}+\mathbf{s}_a)\Longleftrightarrow \widehat{f+g}(\mathbf{0})+\widehat{g}(\mathbf{a})-\widehat{f}(\mathbf{a})=2^m .\]

{\bf Case 2.} Two codewords $\mathbf{c}_1$ and $\mathbf{c}_2$ come from different types, namely,
\[\mathbf{c}_1\in \{\mathbf{s}_a,\mathbf{f}+\mathbf{s}_a,\mathbf{g}+\mathbf{s}_a,\mathbf{f}+\mathbf{g}+\mathbf{s}_a\}\quad \text{and}\quad
\mathbf{c}_2\in \{\mathbf{s}_b,\mathbf{f}+\mathbf{s}_b,\mathbf{g}+\mathbf{s}_b,\mathbf{f}+\mathbf{g}+\mathbf{s}_b\},\]
where both $\mathbf{a}$ and $\mathbf{b}$ are different vectors of ${\mathbb{F}_2^m}$. For convenience, we assume that $\mathbf{a}\neq \mathbf{0}$ and $\mathbf{b}\neq 0$, then we have the following four subcases.

(1) For $\mathbf{c}_1=\mathbf{s}_a$ and $\mathbf{c}_2=\mathbf{s}_b$, since $\mathbf{s}_a$ and $\mathbf{s}_b$ are two different codewords of $S_m$ with $wt(\mathbf{s}_b)=wt(\mathbf{s}_a)=2^{m-1}$. Consequently, $\mathbf{c}_1$ cannot cover $\mathbf{c}_2$, and $\mathbf{c}_2$ cannot cover $\mathbf{c}_1$ as well.

(2) For $\mathbf{c}_1=\mathbf{s}_\alpha$ and $\mathbf{c}_2\in \{\mathbf{f}+\mathbf{s}_\beta,\mathbf{g}+\mathbf{s}_\beta,\mathbf{f}+\mathbf{g}+\mathbf{s}_\beta\}$, where $\{\mathbf{\alpha},\mathbf{\beta}\}=\{\mathbf{a},\mathbf{b}\}$, we only give the proof of $\mathbf{c}_1=\mathbf{s}_a$ and $\mathbf{c}_2=\mathbf{f}+\mathbf{g}+\mathbf{s}_b$, omit the proofs of other cases, whose proofs are similar to this case. Note that $\mathbf{s}_b+\mathbf{s}_a=\mathbf{s}_{a+b}$, by Lemma \ref{lem1} and (\ref{shizi3}), one can obtain
\[\mathbf{c}_1\preceq \mathbf{c}_2\Longleftrightarrow wt(\mathbf{f}+\mathbf{g}+\mathbf{s}_{a+b})=wt(\mathbf{f}+\mathbf{g}+\mathbf{s}_b)-2^{m-1}\Longleftrightarrow \widehat{f+g}(\mathbf{a}+\mathbf{b})-\widehat{f+g}(\mathbf{b})=2^m, \]
and
\[\mathbf{c}_2\preceq \mathbf{c}_1\Longleftrightarrow wt(\mathbf{f}+\mathbf{g}+\mathbf{s}_{a+b})=2^{m-1}-wt(\mathbf{f}+\mathbf{g}+\mathbf{s}_b)\Longleftrightarrow \widehat{f+g}(\mathbf{a}+\mathbf{b})+\widehat{f+g}(\mathbf{b})=2^m .\]

(3) For $\mathbf{c}_1=\mathbf{f}_1+\mathbf{s}_a$ and $\mathbf{c}_2=\mathbf{f}_1+\mathbf{s}_b$, where $\mathbf{f}_1\in\{\mathbf{f},\mathbf{g},\mathbf{f+g}\}$, we only give the proof of $\mathbf{c}_1=\mathbf{f}+\mathbf{s}_a$ and $\mathbf{c}_2=\mathbf{f}+\mathbf{s}_b$, omit the proofs of other cases, whose proofs are similar to this case. Note that $\mathbf{s}_b+\mathbf{s}_a=\mathbf{s}_{a+b}\in S_m$, by Lemma \ref{lem1} and (\ref{shizi3}), one can obtain
\[\mathbf{c}_1\preceq \mathbf{c}_2\Longleftrightarrow wt(\mathbf{s}_{a+b})=wt(\mathbf{f}+\mathbf{s}_b)-wt(\mathbf{f}+\mathbf{s}_a)\Longleftrightarrow \widehat{f}(\mathbf{a})-\widehat{f}(\mathbf{b})=2^m ,\]
and
\[\mathbf{c}_2\preceq \mathbf{c}_1\Longleftrightarrow wt(\mathbf{s}_{a+b})=wt(\mathbf{f}+\mathbf{s}_a)-wt(\mathbf{f}+\mathbf{s}_b)\Longleftrightarrow \widehat{f}(\mathbf{b})-\widehat{f}(\mathbf{a})=2^m.\]

(4) For $\mathbf{c}_1=\mathbf{f}_1+\mathbf{s}_a$ and $\mathbf{c}_2=\mathbf{f}_2+\mathbf{s}_b$, where both $\mathbf{f}_1$ and $\mathbf{f}_2$ belong to $ \{\mathbf{f},\mathbf{g},\mathbf{f+g}\}$ with $\mathbf{f}_1 \neq \mathbf{f}_2$, we only give the proof of $\mathbf{c}_1=\mathbf{f}+\mathbf{s}_a$ and $\mathbf{c}_2=\mathbf{g}+\mathbf{s}_b$, omit the proofs of other cases, whose proofs are similar to this case. Note that $\mathbf{s}_b+\mathbf{s}_a=\mathbf{s}_{a+b}\in S_m$, by Lemma \ref{lem1} and (\ref{shizi3}), we have
\[\mathbf{c}_1\preceq \mathbf{c}_2\Longleftrightarrow wt(\mathbf{f}+\mathbf{g}+\mathbf{s}_{a+b})=wt(\mathbf{g}+\mathbf{s}_b)-wt(\mathbf{f}+\mathbf{s}_a)\Longleftrightarrow \widehat{f+g}(\mathbf{a}+\mathbf{b})+\widehat{f}(\mathbf{a})-\widehat{g}(\mathbf{b})=2^m ,\]
and
\[\mathbf{c}_2\preceq \mathbf{c}_1\Longleftrightarrow wt(\mathbf{f}+\mathbf{g}+\mathbf{s}_{a+b})=wt(\mathbf{f}+\mathbf{s}_a)-wt(\mathbf{g}+\mathbf{s}_b)\Longleftrightarrow \widehat{f+g}(\mathbf{a}+\mathbf{b})+\widehat{g}(\mathbf{b})-\widehat{f}(\mathbf{a})=2^m. \]

Summarizing the discussions above, Theorem \ref{thm2} is immediate. \hfill$\Box$\\

\section{A Class of Minimal Binary Linear Codes from the Generic Construction}
$ $

In this section, using the general construction (\ref{shizi2}), we shall present a class of minimal binary linear codes with $\frac{w_{min}}{w_{max}}\leq \frac{1}{2}$ from a class of specific Boolean functions.

We always assume that $m$ is a positive even integer, and let $t=\frac{m}{2}$. A partial spread of order $\alpha$ (an $\alpha$-spread) in $\mathbb{F}_2^m$ is a set of $\alpha$ $t$-dimensional subspaces $E_{1},E_{2},\cdots, E_{\alpha}$ of $\mathbb{F}_2^m$ such that $E_i\cap E_j=\{\mathbf{0}\}$ for all $1\leq i<j\leq \alpha$. Obviously, the order of a partial spread is less than or equal to $2^t+1$. It is well-known that partial spreads can be used to construct Bent functions \cite{Ding-Heng-Zhou}. In this sequel, we will employ partial spreads to obtain a class of Boolean functions, which can generate a class of minimal binary linear codes with $\frac{w_{min}}{w_{max}}\leq \frac{1}{2}$ by (\ref{shizi2}).

For $1\leq i \leq 2^t+1$, let $f_i:\mathbb{F}_2^m\longrightarrow \mathbb{F}_2$ be the Boolean functions with support $E_i\setminus \{\mathbf{0}\}$, i.e.,
\begin{eqnarray*}
f_i(\mathbf{x})=\left\{\begin{array}{ll}1, &\text{if}\quad \mathbf{x}\in E_i\setminus \{\mathbf{0}\};\\
 0, &\text{otherwise}. \end{array}\right.
\end{eqnarray*}
For $1\leq s,y\leq 2^t+1$, denote $A=\{i_1,\cdots,i_s\}$, $B=\{j_1,\cdots,j_y\}$, $A\cap B=\{t_1,\cdots, t_z\}$, and the symmetric difference of $A$ and $B$ is given by
\[A\triangle B=[A\setminus \{t_1,\cdots, t_z\}]\bigcup[B\setminus \{t_1,\cdots, t_z\}].\]
Define $f$ and $g$ to be respectively
\[f=\sum_{i\in A}f_i\quad \text{and}\quad g=\sum_{j\in B}f_i,\]
then the weight distribution of $\mathcal{C}_{f,g}$ in (\ref{shizi2}) is given by the following theorem.
\begin{thm}\label{lem2}
Let notations defined as above, if $A\triangle B\neq \varnothing$, then the code $\mathcal{C}_{f,g}$ in (\ref{shizi2}) is a $[2^m-1,m+2]$ binary linear code with the weight distribution given in TABLE 1.
\[\text{\small TABLE 1: the weight distribution of $\mathcal{C}_{f,g}$ in Theorem \ref{lem2}}\]

$$
\scalebox{0.7}{
\begin{tabular}{l l}\hline
Weight $w$ & Multiplicity $A_{w}$   \\\hline
$0$ &$1$ \\
 $s(2^t-1)$&$1$\\
 $y(2^t-1)$&$1$\\
$(s+y-2z)(2^t-1)$&$1$\\
$2^{m-1}$&$2^m-1$\\
 $2^{m-1}-s$&$(2^t+1-s)(2^t-1)$\\
 $2^{m-1}-y$&$(2^t+1-y)(2^t-1)$\\
 $2^{m-1}-(s+y-2z)$&$(2^t+1-s-y+2z)(2^t-1)$\\
 $2^{m-1}+2^t-s$&$s(2^t-1)$\\
 $2^{m-1}+2^t-y$&$y(2^t-1)$\\
 $2^{m-1}+2^t-(s+y-2z)$&$(s+y-2z)(2^t-1)$\\\hline
\end{tabular}}\\
$$
\end{thm}
$\mathbf{Proof.}$ By the definitions of $f$ and $f_i$, it is easily verified that
\begin{eqnarray*}
\widetilde{f}(\mathbf{w})=\left\{\begin{array}{lll}s(2^t-1), &\text{if}\quad \mathbf{w}=\mathbf{0};\\
-s, &\text{if}\quad \mathbf{w}\notin {E_i}^\bot\quad\text{for all}\quad i\in A;\\
 2^t-s, &\text{if}\quad \mathbf{w}\neq \mathbf{0}\quad \text{and}\quad \mathbf{w}\in {E_i}^\bot\quad\text{for some}\quad i\in A, \end{array}\right.
\end{eqnarray*}
where ${E_i}^\bot$ is the dual space of $E_i$. Following Lemma \ref{lem3}, we have
\begin{eqnarray*}
\widehat{f}(\mathbf{w})=\left\{\begin{array}{lll}2^m-2s(2^t-1), &1\quad\text{time};\\
2s, &(2^t+1-s)(2^t-1)\quad\text{times};\\
 -2^{t+1}+2s, &s(2^t-1)\quad\text{times}. \end{array}\right.
\end{eqnarray*}
Note that $f+g=\sum_{i\in A\triangle B}f_i$ and $A\triangle B\neq \varnothing$, then we can obtain similarly
\begin{eqnarray*}
\widehat{g}(\mathbf{w})=\left\{\begin{array}{lll}2^m-2y(2^t-1), &1\quad\text{time};\\
2y, &(2^t+1-y)(2^t-1)\quad\text{times};\\
 -2^{t+1}+2y, &y(2^t-1)\quad\text{times}, \end{array}\right.
\end{eqnarray*}
and
\begin{eqnarray*}
\widehat{f+g}(\mathbf{w})=\left\{\begin{array}{lll}2^m-2(s+y-2z)(2^t-1), &1\quad\text{time};\\
2s+2y-4z, &(2^t+1-s-y+2z)(2^t-1)\quad\text{times};\\
 -2^{t+1}+2s+2y-4z, &(s+y-2z)(2^t-1)\quad\text{times}. \end{array}\right.
\end{eqnarray*}
The desired conclusion follows from Theorem \ref{thm1}.

This completes the proof of Theorem \ref{lem2}. \hfill$\Box$\\

For any $x,y,z$ satisfying $A\triangle B\neq \varnothing$, according to Theorem \ref{thm2}, to prove $C_{f,g}$ in Theorem \ref{lem2} is minimal is difficult due to the complexity of calculations. Thus we present $C_{f,g}$ in Theorem \ref{lem2} is minimal for some special cases by Theorem \ref{thm2}. Before giving our main results, we give some lemmas which will be needed later.

Next, we always assume that $s=y$ and $z=1$, i.e., $A=\{i_1,\cdots,i_s\}$, $B=\{j_1,\cdots,j_s\}$, $A\cap B=\{t_1\}$. Recall that
\begin{eqnarray}\label{shizi4}
f=\sum_{i\in A}f_i,\quad g=\sum_{i\in B}f_i,\quad\text{and}\quad f+g=\sum_{i\in A \triangle B}f_i,
\end{eqnarray}
note that $|A\triangle B|=2s-2\leq 2^t+1$, namely, $s\leq 2^{t-1}+1$. We will prove the two conditions of Theorem \ref{thm2} hold for $C_{f,g}$, where $f$, $g$ and $f+g$ are defined in (\ref{shizi4}), we first present condition (1) of Theorem \ref{thm2} holds.
\begin{lemma}\label{lem4}
Let $m\geq6$ and $s$ be an integer with $2\leq s\leq 2^{t-1}$, $f$, $g$ and $f+g$ be the Boolean functions defined in (\ref{shizi4}), denote $F=\{f,g,f+g\}$. Then for any $\mathbf{h},\textbf{l}\in\mathbb{F}_2^m$ with $\mathbf{h}\neq\textbf{l}$, we have
\[\widehat{f}_1(\mathbf{h})+\widehat{f}_2(\textbf{l})\neq 2^m \quad \text{and} \quad \widehat{f}_1(\mathbf{h})-\widehat{f}_2(\textbf{l})\neq 2^m\]
where $f_1,f_2\in F$.
\end{lemma}
$\mathbf{Proof.}$ From the proof of Theorem \ref{lem2}, we have
\[\text{\small TABLE 2: the values of the Walsh transforms for $f$, $g$ and $f+g$ defined in (\ref{shizi4})}\]
$$
\scalebox{0.7}{
\begin{tabular}{llllll}\hline
the value of $\widehat{f}(\mathbf{w})$& condiions & the value of $\widehat{g}(\mathbf{w})$& conditions &the value of $\widehat{f+g}(\mathbf{w})$ & conditions  \\\hline
$2^m-2s(2^t-1)$ &$\mathbf{w}=\mathbf{0}$&$2^m-2s(2^t-1)$&$\mathbf{w}=\mathbf{0}$&$2^m-(4s-4)(2^t-1)$& $\mathbf{w}=\mathbf{0}$\\
 $2s$&$\mathbf{w}\notin {E_i}^\bot$ for all $i\in A$&2s&$\mathbf{w}\notin {E_i}^\bot$ for all $i\in B$&4s-4&$\mathbf{w}\notin {E_i}^\bot$ for all $i\in A\triangle B$\\
 $-2^{t+1}+2s$&$\mathbf{w}\in {E_i}^\bot$ for some $i\in A$&$-2^{t+1}+2s$&$\mathbf{w}\in {E_i}^\bot$ for some $i\in B$&$-2^{t+1}+4s-4$&$\mathbf{w}\in {E_i}^\bot$ for some $i\in A\triangle B$\\\hline
\end{tabular}}\\
$$
According to different choices of $f_1$ and $f_2$, we divide into the following six cases to prove Lemma \ref{lem4}.

{\bf Case 1}. If $f_1=f_2=f$, there are five subcases according to choices of both $\mathbf{h}$ and $\textbf{l}$. We only give the proofs of two cases as follows, and omit the proofs of other cases, whose proofs are similar.

(1) If $\mathbf{h}=\mathbf{0}$ and $\textbf{l}\notin {E_i}^\bot$ for all $i\in A$, since $m=2t$ and $t\geq 3$, we have
\[\widehat{f}(\mathbf{0})+\widehat{f}(\textbf{l})\neq 2^m\Leftrightarrow2^m-2s(2^t-1)+2s\neq 2^m\Leftrightarrow2s\neq2s(2^t-1)(\text{obviously}),\]
\[\widehat{f}(\mathbf{0})-\widehat{f}(\textbf{l})\neq 2^m\Leftrightarrow2^m-2s(2^t-1)-2s\neq 2^m\Leftrightarrow2s\cdot2^t\neq 0(\text{obviously}),\]
and
\[\widehat{f}(\textbf{l})-\widehat{f}(\textbf{0})\neq 2^m\Leftrightarrow2s-2^m+2s(2^t-1)\neq 2^m\Leftrightarrow2s\cdot2^t\neq 2^{m+1}\Leftrightarrow s\neq 2^t.\]

(2) If $\mathbf{h}=\mathbf{0}$ and $\textbf{l}\in {E_i}^\bot$ for some $i\in A$, since $m=2t$ and $t\geq 3$, we have
\[\widehat{f}(\mathbf{0})+\widehat{f}(\textbf{l})\neq 2^m\Leftrightarrow2^m-2s(2^t-1)-2^{t+1}+2s\neq 2^m\Leftrightarrow-2s(2^t-2)-2^{t+1}\neq 0(\text{obviously}),\]
\[\widehat{f}(\mathbf{0})-\widehat{f}(\textbf{l})\neq 2^m\Leftrightarrow2^m-2s(2^t-1)+2^{t+1}-2s\neq 2^m\Leftrightarrow2s\cdot2^t\neq 2^{t+1}\Leftrightarrow s\neq1,\]
and
\[\widehat{f}(\textbf{l})-\widehat{f}(\textbf{0})\neq 2^m\Leftrightarrow-2^{t+1}+2s-2^m+2s(2^t-1)\neq 2^m\Leftrightarrow2s\cdot2^t\neq 2^{m+1}+2^{t+1}\Leftrightarrow s\neq 2^t+1.\]

{\bf Case 2}. If $f_1=f_2=g$, since the proof of this case is similar to that of {\bf Case 1}, we omit it.

{\bf Case 3}. If $f_1=f_2=f+g$, there are five subcases according to choices of both $\mathbf{h}$ and $\textbf{l}$. We only give the proof of $\mathbf{h}=\mathbf{0}$ and $\textbf{l}\notin {E_i}^\bot$ for all $i\in A\triangle B$, and omit the proofs of other cases, whose proofs are similar.

(1) If $\mathbf{h}=\mathbf{0}$ and $\textbf{l}\notin {E_i}^\bot$ for all $i\in A\triangle B$, since $m=2t$ and $t\geq 3$, we have
\[\widehat{f+g}(\mathbf{0})+\widehat{f+g}(\textbf{l})\neq 2^m\Leftrightarrow2^m-(4s-4)(2^t-1)+4s-4\neq 2^m\Leftrightarrow4s-4\neq(4s-4)(2^t-1)\Leftrightarrow s\neq1,\]
\[\widehat{f+g}(\mathbf{0})-\widehat{f+g}(\textbf{l})\neq 2^m\Leftrightarrow2^m-(4s-4)(2^t-1)-4s+4\neq 2^m\Leftrightarrow(4s-4)2^t\neq0\Leftrightarrow s\neq1,\]
and
\[\widehat{f+g}(\textbf{l})-\widehat{f+g}(\textbf{0})\neq 2^m\Leftrightarrow4s-4-2^m+(4s-4)(2^t-1)\neq 2^m\Leftrightarrow(4s-4)2^t\neq 2^{m+1}\Leftrightarrow s\neq 2^{t-1}+1.\]

{\bf Case 4}. If $f_1=f$ and $f_2=g$, since the proof of this case is similar to that of {\bf Case 1}, we omit it.

{\bf Case 5}. If $f_1=f+g$ and $f_2=f$, there are five subcases according to choices of both $\mathbf{h}$ and $\textbf{l}$, we only give the proof of $\mathbf{h}=\mathbf{0}$ and $\textbf{l}\notin {E_i}^\bot$ for all $i\in A$, and omit the proofs of other cases, whose proofs are similar.

(1) If $\mathbf{h}=\mathbf{0}$ and $\textbf{l}\notin {E_i}^\bot$ for all $i\in A$, since $m=2t$, $t\geq 3$ and $2\leq s\leq 2^{t-1}$, we have
\[\widehat{f+g}(\mathbf{0})+\widehat{f}(\textbf{l})\neq 2^m\Leftrightarrow2^m-(4s-4)(2^t-1)+2s\neq 2^m\Leftrightarrow(4s-4)(2^t-1)\neq 2s(\text{obviously}),\]
\[\widehat{f+g}(\mathbf{0})-\widehat{f}(\textbf{l})\neq 2^m\Leftrightarrow2^m-(4s-4)(2^t-1)-2s\neq 2^m\Leftrightarrow-(4s-4)(2^t-1)-2s\neq0(\text{obviously}),\]
and
\[\widehat{f}(\textbf{l})-\widehat{f+g}(\textbf{0})\neq 2^m\Leftrightarrow2s-2^m+(4s-4)(2^t-1)\neq 2^m\Leftrightarrow s+(2s-2)(2^t-1)<2^m-2^{t-1}\neq 2^{m}.\]

{\bf Case 6}. If $f_1=f+g$ and $f_2=g$, since the proof of this case is similar to that of {\bf Case 5}, we omit it.

Summarizing the discussions above, we complete the proof of Lemma \ref{lem4}. \hfill$\Box$\\

The following three lemmas will be needed to prove Theorem \ref{thm2} (2).
\begin{lemma}\label{lem5}
Let $m\geq6$ and $s$ be an integer with $2 \leq s\leq 2^{t-1}$, $f$, $g$ and $f+g$ be the Boolean functions defined in (\ref{shizi4}), denote $F=\{f,g,f+g\}$. Then for any $\mathbf{h} \in\mathbb{F}_2^m$, we have
\[\widehat{f}_1(\mathbf{0})+\widehat{f}_2(\mathbf{h})-\widehat{f_1+f_2}(\mathbf{h})\neq 2^m,\]
where $f_1,f_2\in F$ with $f_1\neq f_2$.
\end{lemma}
$\mathbf{Proof.}$ According to different choices of $f_1$ and $f_2$, we divide into the following six cases to prove Lemma \ref{lem5}.

{\bf Case 1}. If $f_1=f$ and $f_2=g$, according to the choice of $\mathbf{h}$ by TABLE 2, we will divide this case into three subcases.

(1) If $\mathbf{h}=\mathbf{0}$, since $m=2t$ and $t\geq3$, then $\widehat{f}(\mathbf{0})+\widehat{g}(\mathbf{0})-\widehat{f+g}(\mathbf{0})\neq 2^m$, i.e.,
\[2^m-2s(2^t-1)+2^m-2s(2^t-1)-2^m+(4s-4)(2^t-1)\neq 2^m\Leftrightarrow(4s-4)(2^t-1)\neq 4s(2^t-1)(\text{obviously}).\]

(2) If $\mathbf{h}\notin {E_i}^\bot$ for all $i\in A\triangle B\cup\{z_1\}$, since $m=2t$ and $t\geq3$, then $\widehat{f}(\mathbf{0})+\widehat{g}(\mathbf{h})-\widehat{f+g}(\mathbf{h})\neq 2^m$, i.e.,
\[2^m-2s(2^t-1)+2s-4s+4\neq 2^m\Leftrightarrow 2s\cdot2^t\neq 4(\text{obviously}).\]

(3) If $\mathbf{h}\in {E_i}^\bot$ for some $i\in A\triangle B\cup\{z_1\}$, we divide this case into three subcases, namely, $i=z_1$, $i\in A\setminus\{z_1\}$, and $i\in B\setminus\{z_1\}$. Since $m=2t$ and $t\geq3$, then for $i=z_1$, we have
\[\widehat{f}(\mathbf{0})+\widehat{g}(\mathbf{h})-\widehat{f+g}(\mathbf{h})\neq 2^m\Leftrightarrow 2^m-2s(2^t-1)-2^{t+1}+2s-4s+4\neq 2^m\Leftrightarrow-2s\cdot2^t-2^{t+1}+4\neq 0(\text{obviously});\]
for $i\in A\setminus\{z_1\}$, we have
\[\widehat{f}(\mathbf{0})+\widehat{g}(\mathbf{h})-\widehat{f+g}(\mathbf{h})\neq 2^m\Leftrightarrow 2^m-2s(2^t-1)+2s+2^{t+1}-4s+4\neq 2^m\Leftrightarrow s\cdot2^t\neq 2^t+2(\text{obviously});\]
and for $i\in B\setminus\{z_1\}$, we have
\[\widehat{f}(\mathbf{0})+\widehat{g}(\mathbf{h})-\widehat{f+g}(\mathbf{h})\neq 2^m\Leftrightarrow2^m-2s(2^t-1)-2^{t+1}+2s+2^{t+1}-4s+4\neq 2^m\Leftrightarrow s(2^t-1)\neq2-s(\text{obviously}).\]

{\bf Case 2}. If $f_1=g$ and $f_2=f$, since the proof is similar to that of {\bf Case 1}, we omit it.

{\bf Case 3}. If $f_1=f$ and $f_2=f+g$, according to the choice of $\mathbf{h}$ by TABLE 2, we will divide this case into three subcases.

(1) If $\mathbf{h}=\mathbf{0}$, since $m=2t$ and $t\geq3$, then $\widehat{f}(\mathbf{0})+\widehat{f+g}(\mathbf{0})-\widehat{g}(\mathbf{0})\neq 2^m$, i.e.,
\[2^m-2s(2^t-1)+2^m-(4s-4)(2^t-1)-2^m+2s(2^t-1)\neq 2^m\Leftrightarrow(4s-4)(2^t-1)\neq0(\text{obviously}).\]

(2) If $\mathbf{h}\notin {E_i}^\bot$ for all $i\in A\triangle B\cup\{z_1\}$, since $m=2t$ and $t\geq3$, then $\widehat{f}(\mathbf{0})+\widehat{f+g}(\mathbf{h})-\widehat{g}(\mathbf{h})\neq 2^m$, i.e.,
\[2^m-2s(2^t-1)+4s-4-2s\neq 2^m\Leftrightarrow s(2^t-1)\neq s-2(\text{obviously}).\]

(3) If $\mathbf{h}\in {E_i}^\bot$ for some $i\in A\triangle B\cup\{z_1\}$, we divide this case into three subcases, namely, $i=z_1$, $i\in A\setminus\{z_1\}$, and $i\in B\setminus\{z_1\}$. Since $m=2t$ and $t\geq3$, for $i=z_1$, we have
\[\widehat{f}(\mathbf{0})+\widehat{f+g}(\mathbf{h})-\widehat{g}(\mathbf{h})\neq 2^m\Leftrightarrow2^m-2s(2^t-1)+4s-4+2^{t+1}-2s\neq 2^m\Leftrightarrow s(2^t-2)\neq 2^t-2(\text{obviously});\]
for $i\in A\setminus\{z_1\}$, we have
\[\widehat{f}(\mathbf{0})+\widehat{f+g}(\mathbf{h})-\widehat{g}(\mathbf{h})\neq 2^m\Leftrightarrow2^m-2s(2^t-1)-2^{t+1}+4s-4-2s\neq 2^m\Leftrightarrow -s(2^t-2)-(2^t+2)\neq 0(\text{obviously});\]
and for $i\in B\setminus\{z_1\}$, we have
\[\widehat{f}(\mathbf{0})+\widehat{f+g}(\mathbf{h})-\widehat{g}(\mathbf{h})\neq 2^m\Leftrightarrow2^m-2s(2^t-1)-2^{t+1}+4s-4+2^{t+1}-2s\neq 2^m\Leftrightarrow s(2^t-1)\neq s-2(\text{obviously}).\]

{\bf Case 4}. If $f_1=g$ and $f_2=f+g$, since the proof is similar to that of {\bf Case 3}, we omit it.

{\bf Case 5}. If $f_1=f+g$ and $f_2=f$, according to the choice of $\mathbf{h}$ by TABLE 2, we will divide this case into three subcases.

(1) If $\mathbf{h}=\mathbf{0}$, since the proof is similar to that of {\bf Case 3} (1), we omit it.

(2) If $\mathbf{h}\notin {E_i}^\bot$ for all $i\in A\triangle B\cup\{z_1\}$, since $m=2t$ and $t\geq3$, then $\widehat{f+g}(\mathbf{0})+\widehat{f}(\mathbf{h})-\widehat{g}(\mathbf{h})\neq 2^m$, i.e.,
\[2^m-(4s-4)(2^t-1)+2s-2s\neq 2^m\Leftrightarrow (4s-4)(2^t-1)\neq 0(\text{obviously}).\]

(3) If $\mathbf{h}\in {E_i}^\bot$ for some $i\in A\triangle B\cup\{z_1\}$, we divide this case into three subcases, namely, $i=z_1$, $i\in A\setminus\{z_1\}$, and $i\in B\setminus\{z_1\}$. Since $m=2t$ and $t\geq3$, for $i=z_1$, we have
\[\widehat{f+g}(\mathbf{0})+\widehat{f}(\mathbf{h})-\widehat{g}(\mathbf{h})\neq 2^m\Leftrightarrow2^m-(4s-4)(2^t-1)-2^{t+1}+2s+2^{t+1}-2s\neq 2^m\Leftrightarrow (4s-4)(2^t-1)\neq 0(\text{obviously});\]
for $i\in A\setminus\{z_1\}$, we have
\[\widehat{f+g}(\mathbf{0})+\widehat{f}(\mathbf{h})-\widehat{g}(\mathbf{h})\neq 2^m\Leftrightarrow2^m-(4s-4)(2^t-1)-2^{t+1}+2s-2s\neq 2^m\Leftrightarrow -(4s-4)(2^t-1)-2^{t+1}\neq 0(\text{obviously});\]
and for $i\in B\setminus\{z_1\}$, we have
\[\widehat{f+g}(\mathbf{0})+\widehat{f}(\mathbf{h})-\widehat{g}(\mathbf{h})\neq 2^m\Leftrightarrow2^m-(4s-4)(2^t-1)+2s+2^{t+1}-2s\neq 2^m\Leftrightarrow (2s-2)(2^t-1)\neq 2^t (\text{obviously}).\]

{\bf Case 6}. If $f_1=f+g$ and $f_2=g$, since the proof is similar to that of {\bf Case 5}, we omit it.

Summarizing the discussions above, we complete the proof of Lemma \ref{lem5}. \hfill$\Box$\\

\begin{lemma}\label{lem7}
Let $m\geq6$ and $s$ be an integer with $2 \leq s\leq 2^{t-1}$, $f$, $g$ and $f+g$ be the Boolean functions defined in (\ref{shizi4}), denote $F=\{f,g,f+g\}$. Then for any $\mathbf{h} \in\mathbb{F}_2^m\setminus\{\mathbf{0}\}$, we have
\[\widehat{f}_1(\mathbf{h})+\widehat{f}_2(\mathbf{h})-\widehat{f_1+f_2}(\mathbf{0})\neq 2^m,\quad \text{and}\quad \widehat{f}_1(\mathbf{h})+\widehat{f}_2(\mathbf{0})-\widehat{f_1+f_2}(\mathbf{h})\neq 2^m \]
where $f_1,f_2\in F$ with $f_1\neq f_2$.
\end{lemma}
$\mathbf{Proof.}$ According to the different choices of $f_1$ and $f_2$, we have the following six cases.

{\bf Case 1}. If $f_1=f$ and $f_2=g$, according to the choice of $\mathbf{h}$ by TABLE 2, we will divide this case into two subcases.

(1) If $\mathbf{h}\notin {E_i}^\bot$ for all $i\in A\triangle B\cup\{z_1\}$, since $m=2t$ and $t\geq3$, we have
\[\widehat{f}(\mathbf{h})+\widehat{g}(\mathbf{h})-\widehat{f+g}(\mathbf{0})\neq 2^m\Leftrightarrow 2s+2s-2^m+(4s-4)(2^t-1)\neq 2^m\Leftrightarrow(2s-2)2^t\neq 2^m-2(\text{obviously}),\]
and
\[\widehat{f}(\mathbf{h})+\widehat{g}(\mathbf{0})-\widehat{f+g}(\mathbf{h})\neq 2^m\Leftrightarrow 2s+2^m-2s(2^t-1)-4s+4\neq 2^m\Leftrightarrow 2s\cdot 2^t\neq 4 (\text{obviously}).\]

(2) If $\mathbf{h}\in {E_i}^\bot$ for some $i\in A\triangle B\cup\{z_1\}$, we divide this case into three subcases, namely, $i=z_1$, $i\in A\setminus\{z_1\}$, and $i\in B\setminus\{z_1\}$. Since $m=2t$ and $t\geq3$, for $i=z_1$, we have
\[\widehat{f}(\mathbf{h})+\widehat{g}(\mathbf{h})-\widehat{f+g}(\mathbf{0})\neq 2^m\Leftrightarrow2(-2^{t+1}+2s)-2^m+(4s-4)(2^t-1)\neq 2^m\Leftrightarrow 2s\cdot2^t-2^{t+2}+2\leq 2^m-2^{t+2}+2\neq 2^m;\]
and
\[\widehat{f}(\mathbf{h})+\widehat{g}(\mathbf{0})-\widehat{f+g}(\mathbf{h})\neq 2^m\Leftrightarrow -2^{t+1}+2s+2^m-2s(2^t-1)-4s+4\neq 2^m\Leftrightarrow s\cdot2^t\neq 4-2^{t+1}(\text{obviously}).\]
And for $i\in A\setminus\{z_1\}$, we have
\[\widehat{f}(\mathbf{h})+\widehat{g}(\mathbf{h})-\widehat{f+g}(\mathbf{0})\neq 2^m\Leftrightarrow -2^{t+1}+2s+2s-2^m+(4s-4)(2^t-1)\neq 2^m\Leftrightarrow (s-1)2^t\neq 2^{m-1}+2^{t-1}-1(\text{obvious});\]
and
\[\widehat{f}(\mathbf{h})+\widehat{g}(\mathbf{0})-\widehat{f+g}(\mathbf{h})\neq 2^m\Leftrightarrow -2^{t+1}+2s+2^m-2s(2^t-1)+2^{t+1}-4s+4\neq 2^m\Leftrightarrow s\cdot2^t\neq 2(\text{obvious}).\]
For $i\in B\setminus\{z_1\}$, the proof is similar to that of $i\in A\setminus\{z_1\}$, we omit it.

{\bf Case 2}. If $f_1=g$ and $f_2=f$, since the proof is similar to that of {\bf Case 1}, we omit it.

{\bf Case 3}. If $f_1=f$ and $f_2=f+g$, according to the choice of $\mathbf{h}$ by TABLE 2, we will divide this case into two subcases.

(1) If $\mathbf{h}\notin {E_i}^\bot$ for all $i\in A\triangle B\cup\{z_1\}$, since $m=2t$ and $t\geq3$, we have
\[\widehat{f}(\mathbf{h})+\widehat{f+g}(\mathbf{h})-\widehat{g}(\mathbf{0})\neq 2^m\Leftrightarrow 2s+4s-4-2^m+2s(2^t-1)\neq 2^m\Leftrightarrow2s(2^t+2)\neq 2^m+2(\text{obviously}),\]
and
\[\widehat{f}(\mathbf{h})+\widehat{f+g}(\mathbf{0})-\widehat{g}(\mathbf{h})\neq 2^m\Leftrightarrow 2s+2^m-(4s-4)(2^t-1)-2s\neq 2^m\Leftrightarrow -(4s-4)(2^t-1)\neq0 (\text{obviously}).\]

(2) If $\mathbf{h}\in {E_i}^\bot$ for some $i\in A\triangle B\cup\{z_1\}$, we divide this case into three subcases, namely, $i=z_1$, $i\in A\setminus\{z_1\}$, and $i\in B\setminus\{z_1\}$. Since $m=2t$ and $t\geq3$, for $i=z_1$, we have
\[\widehat{f}(\mathbf{h})+\widehat{f+g}(\mathbf{h})-\widehat{g}(\mathbf{0})\neq 2^m\Leftrightarrow-2^{t+1}+2s+4s-4-2^m+2s(2^t-1)\neq 2^m\Leftrightarrow s\neq 1+\frac{m}{2^t+2}(\text{obviously});\]
and
\[\widehat{f}(\mathbf{h})+\widehat{f+g}(\mathbf{0})-\widehat{g}(\mathbf{h})\neq 2^m\Leftrightarrow -2^{t+1}+2s+2^m-(4s-4)(2^t-1)+2^{t+1}-2s\neq 2^m\Leftrightarrow -(4s-4)(2^t-1)\neq 0(\text{obviously}).\]
For $i\in A\setminus\{z_1\}$, we have
\[\widehat{f}(\mathbf{h})+\widehat{f+g}(\mathbf{h})-\widehat{g}(\mathbf{0})\neq 2^m\Leftrightarrow -2^{t+1}+2s-2^{t+1}+4s-4-2^m+2s(2^t-1)\neq 2^m\Leftrightarrow s\neq 2^t+\frac{2}{2^t+2}(\text{obviously});\]
and
\[\widehat{f}(\mathbf{h})+\widehat{f+g}(\mathbf{0})-\widehat{g}(\mathbf{h})\neq 2^m\Leftrightarrow -2^{t+1}+2s+2^m-(4s-4)(2^t-1)-2s\neq 2^m\Leftrightarrow-2^{t-1}-(s-1)(2^t-1)\neq 0(\text{obviously}).\]
And for $i\in B\setminus\{z_1\}$, we have
\[\widehat{f}(\mathbf{h})+\widehat{f+g}(\mathbf{h})-\widehat{g}(\mathbf{0})\neq 2^m\Leftrightarrow 2s-2^{t+1}+4s-4-2^m+2s(2^t-1)\neq 2^m\Leftrightarrow s\neq 1+\frac{m}{2^t+2}(\text{obviously});\]
and
\[\widehat{f}(\mathbf{h})+\widehat{f+g}(\mathbf{0})-\widehat{g}(\mathbf{h})\neq 2^m\Leftrightarrow 2s+2^m-(4s-4)(2^t-1)+2^{t+1}-2s\neq 2^m\Leftrightarrow(2s-2)(2^t-1)\neq 2^t(\text{obviously}).\]

If $f_1=g$ and $f_2=f+g$, or both $f_1=f+g$ and $f_2=f$, or both $f_1=f+g$ and $f_2=g$, since their proofs is similar to that of {\bf Case 3}, we omit them.

Summarizing the discussions above, we complete the proof of Lemma \ref{lem7}. \hfill$\Box$\\

\begin{lemma}\label{lem8}
Let $m\geq6$ and $s$ be an integer with $2\leq s\leq 2^{t-1}$, $f$, $g$ and $f+g$ be the Boolean functions defined in (\ref{shizi4}), denote $F=\{f,g,f+g\}$. Then for any $\mathbf{h}, \textbf{l}\in\mathbb{F}_2^m\setminus\{\mathbf{0}\}$ with $\mathbf{h}\neq \textbf{l}$, we have
\[\widehat{f}_1(\mathbf{h}+\textbf{l})+\widehat{f}_2(\mathbf{h})-\widehat{f_1+f_2}(\textbf{l})\neq 2^m\quad \text{and}\quad \widehat{f}_1(\mathbf{h}+\textbf{l})+\widehat{f}_2(\textbf{l})-\widehat{f_1+f_2}(\mathbf{h})\neq 2^m, \]
where $f_1,f_2\in F$ with $f_1\neq f_2$.
\end{lemma}
$\mathbf{Proof.}$ Since $\mathbf{h}, \mathbf{l}$ are nonzero vectors with $\mathbf{h}\neq \mathbf{l}$, by TABLE 2, we have $\widehat{f}_i(\mathbf{a})\in \{2s,-2^{t+1}+2s\}$ and $\widehat{f+g}(\mathbf{a})\in \{4s-4,-2^{t+1}+4s-4\}$, where $f_i\in\{f,g\}$ and $\mathbf{a}\in\{\mathbf{h}, \mathbf{l},\mathbf{h}+\mathbf{l}\}$. By $s\leq 2^{t-1}$, we know that $-2^{t+1}+2s\leq 0$ and $-2^{t+1}+4s-4<0$, thus we have
\[|-2^{t+1}+2s|<2^{t+1},\quad |-2^{t+1}+4s-4|<2^{t+1},\quad  |2s|<2^{t+1}, \quad \text{and}\quad |4s-4|\leq 2^{t+1}.\]
And so, by $m=2t$ and $t\geq3$, for any $f_1,f_2\in\{f,g,f+g\}$ with $f_1\neq f_2$, one can obtain
\begin{eqnarray*}
\widehat{f}_1(\mathbf{h}+\textbf{l})+\widehat{f}_2(\mathbf{h})-\widehat{f_1+f_2}(\textbf{l})&\leq&|\widehat{f}_1(\mathbf{h}+\textbf{l})|
+|\widehat{f}_2(\mathbf{h})|+|\widehat{f_1+f_2}(\textbf{l})|\\
&<&32^{t+1}<2^{t+2}<2^{m},
\end{eqnarray*}
and
\begin{eqnarray*}
\widehat{f}_1(\mathbf{h}+\textbf{l})+\widehat{f}_2(\mathbf{l})-\widehat{f_1+f_2}(\textbf{h})&\leq&|\widehat{f}_1(\mathbf{h}+\textbf{l})|
+|\widehat{f}_2(\mathbf{l})|+|\widehat{f_1+f_2}(\textbf{h})|\\
&<&32^{t+1}<2^{t+2}<2^{m}.
\end{eqnarray*}

This completes the proof of Lemma \ref{lem8}. \hfill$\Box$\\

With the preparations above, we have the following theorem.

\begin{thm}\label{thm3}
Let $s=y$ and $z=1$, if $m\geq 6$ and $2\leq s\leq 2^{t-1}-1$, then the code $\mathcal{C}_{f,g}$ in Theorem \ref{lem2} is a $[2^m-1,m+2]$ binary minimal linear code, and the weight distribution is given by TABLE 3. Furthermore, if $s\leq 2^{t-2}$, then $\frac{w_{min}}{w_{max}}\leq \frac{1}{2}$.
\[\text{\small TABLE 3: the weight distribution of $\mathcal{C}_{f,g}$ in Theorem \ref{thm3}}\]

$$
\scalebox{0.7}{
\begin{tabular}{l l}\hline
Weight $w$ & Multiplicity $A_{w}$   \\\hline
$0$ &$1$ \\
 $s(2^t-1)$&$2$\\
$(2s-2)(2^t-1)$&$1$\\
$2^{m-1}$&$2^m-1$\\
 $2^{m-1}-s$&$2(2^t+1-s)(2^t-1)$\\
 $2^{m-1}-(2s-2)$&$(2^t+1-2s+2)(2^t-1)$\\
 $2^{m-1}+2^t-s$&$2s(2^t-1)$\\
 $2^{m-1}+2^t-(2s-2)$&$(2s-2)(2^t-1)$\\\hline
\end{tabular}}\\
$$
\end{thm}
$\mathbf{Proof.}$ We first show the code $\mathcal{C}_{f,g}$ in Theorem \ref{thm3} is minimal by Theorem \ref{thm2}. In fact, by lemma \ref{lem4}, Theorem \ref{thm2} (1) is true for $\mathcal{C}_{f,g}$. We will divide into three cases to present Theorem \ref{thm2} (2) is true for $\mathcal{C}_{f,g}$.

{\bf Case 1}. If $\mathbf{h}=\textbf{l}\in \mathbb{F}_2^m$, Theorem \ref{thm2} (2) is true for $\mathcal{C}_{f,g}$ by Lemma \ref{lem5}.

{\bf Case 2}. If one of $\mathbf{h}$ and $\textbf{l}$ is the zero vector, Theorem \ref{thm2} (2) is true for $\mathcal{C}_{f,g}$ by Lemma \ref{lem7}.

{\bf Case 3}. If $\mathbf{h}$ and $\textbf{l}$ are different nonzero vectors, Theorem \ref{thm2} (2) is true for $\mathcal{C}_{f,g}$ by Lemma \ref{lem8}.

When $s=y$ and $z=1$, the weight distribution of $\mathcal{C}_{f,g}$ follows from Theorem \ref{lem2}. Since $s\leq 2^{t-1}-1$, then $w_{min}$ of $\mathcal{C}_{f,g}$ is $s(2^t-1)$, and $w_{max}$ of $\mathcal{C}_{f,g}$ is $2^{m-1}+2^t-s$. It is easy to verify that $\frac{w_{min}}{w_{max}}\leq \frac{1}{2}$ by $s\leq 2^{t-2}$.

This completes the proof of Theorem \ref{thm3}. \hfill$\Box$\\

\begin{rem}
Note that the code length and the minimum distance of $\mathcal{C}_{f,g}$ in Theorem \ref{thm3} is the same as those of the code in Theorem 18 \cite{Ding-Heng-Zhou}. The dimension of the code in Theorem 18 is $m+1$, but the dimension of $\mathcal{C}_{f,g}$ is $m+2$, which is bigger than the dimension of the code in Theorem 18.
\end{rem}

Based on Magma's program, the following two examples are presented, which is accordant with Theorem \ref{thm3}.
\begin{exmp}
(1) Let $m=6$ and $s=2$, then the code $\mathcal{C}_{f,g}$ in Theorem \ref{thm3} is minimal with parameters $[63,8,14]$ and the weight enumerator is
\[1+3z^{14}+147z^{30}+63z^{32}+42z^{38}.\]
Clearly, $\frac{w_{min}}{w_{max}}=\frac{14}{38}< \frac{1}{2}$.

(2) Let $m=8$ and $s=4$, then the code $\mathcal{C}_{f,g}$ in Theorem \ref{thm3} is minimal with parameters $[255,10,60]$ and the weight enumerator is
\[1+2z^{60}+z^{90}+165z^{122}+390z^{124}+255z^{128}+90z^{138}+120z^{140}.\]
Obviously, $\frac{w_{min}}{w_{max}}=\frac{60}{140}< \frac{1}{2}$.
\end{exmp}

\section{Conclusions and Further Study}
$ $

In this paper, we present a generic construction for binary linear codes with dimension $m+2$, and give a necessary and sufficient condition for this binary linear code to be minimal. And then, based on this condition and exponential sums, a new class of minimal binary linear codes violating the Ashikhmin-Barg condition are constructed.

Based on the generic construction (\ref{shizi2}), it would be interesting to find more minimal binary linear codes violating the Ashikhmin-Barg condition. Another interesting research challenge is to extend the construction (\ref{shizi2}) to nonbinary alphabet to obtain minimal linear codes violating the Ashikhmin-Barg condition.

 \end{document}